\documentclass[12pt]{iopart}
\usepackage{graphicx}
\newcommand{\beq}{\begin{equation}}
\newcommand{\eeq}{\end{equation}}
\newcommand{\beqa}{\begin{eqnarray}}
\newcommand{\eeqa}{\end{eqnarray}}
\newcommand{\ba}{\begin{array}}
\newcommand{\ea}{\end{array}} 

\begin{document} 

\title[Dimensional Effects on Solitons with Normal and Anomalous Dispersion]
{Dimensional Effects on Solitonic Matter and Optical Waves
with Normal and Anomalous Dispersion} 
\author{L. Salasnich$^{1}$, A. Parola$^{2}$ and L. Reatto$^{3}$} 
\address{$^{1}$CNR-INFM and CNISM, Unit\`a di Milano, \\ 
Via Celoria 16, 20133 Milano, Italy \\ 
$^{2}$Dipartimento di Fisica e Matematica, Universit\`a dell'Insubria, \\ 
Via Valeggio 11, 22100 Como, Italy\\
$^{3}$Dipartimento di Fisica and CNISM, Universit\`a di Milano \\
Via Celoria 16, 20133 Milano, Italy}

\begin{abstract} 
We investigate bright and dark solitons with anomalous or normal 
dispersion and under transverse harmonic confinement. 
In matter waves, positive atomic mass implies 
anomalous dispersion (kinetic spreading) while negative mass 
gives normal dispersion (kinetic shrinking). 
We find that, contrary to the strictly one-dimensional case, 
the axial and transverse profiles of these solitons 
crucially depend on the strength of the nonlinearity 
and on their dispersive properties. 
In particular, we show that, like bright solitons with anomalous 
dispersion, also dark solitons with normal dispersion 
disappear at a critical axial density. Our predictions are useful 
for the study of atomic matter waves in 
Bose-Einstein condensates and also for optical bullets 
in inhomogeneous Kerr media. 
\end{abstract} 


\section{Introduction} 

Matter waves made of Bose condensed alkali-metal atoms \cite{r1} and 
optical waves in nonlinear Kerr media \cite{r2} are 
accurately described by a three-dimensional 
nonlinear Schr\"odinger equation (3D NLSE). In fact, 
the NLSE is a unifying theoretical tool of nonlinear 
optics and of the new field of research 
called nonlinear atom optics \cite{r3,r4}. 
Many experiments have been devoted to the study of solitary 
waves in both matter waves \cite{r5,r6,r7,r8,r9} 
and optical waves \cite{r10}. 
Bright and dark solitons are usually analyzed in 
quasi-1D configurations, where analytical solutions 
of the 1D cubic NLSE are available \cite{r11,r12}. 
On the other hand, it has been shown 
that dimensional effects can be quite important: 
under transverse confinement atomic bright solitons with 
anomalous dispersion, i.e. kinetic spreading, 
become unstable if the transverse confinement 
is not sufficiently strong \cite{r13}. 
\par 
In this paper we analyze bright and dark solitons with anomalous 
and normal dispersion (kinetic shrinking) 
taking into account the transverse 
width of the solitary wave. We introduce a non-polynomial 
Schr\"odinger equation, which extends that 
we derived some years ago \cite{r14} by including the case of 
the normal dispersion. In this way, we find new 
analytical solutions for 3D bright solitons under 
transverse harmonic confinement and show that, 
contrary to the strictly 1D case, 3D black and gray 
solitons with normal dispersion exist only below 
a critical axial density. Our predictions can be tested 
not only with matter waves in a periodic optical potential 
along the axial direction \cite{r15} 
but also with optical bullets in a inhomogeneous 
graded-index Kerr medium \cite{r16}. 

\section{Matter waves and optical pulses} 

The 3D cubic nonlinear Schr\"odinger equation (3D CNLSE), 
which describes both matter waves in a Bose-Einstein condensate 
and optical pulses in a inhomogeneous nonlinear Kerr medium, 
is given by 
\beq 
i{\partial \psi \over \partial t} = 
\Big[ 
- {1\over 2} \nabla^2_{\bot} 
- {\delta_D \over 2} {\partial^2 \over \partial z^2} 
+ {1\over 2} (x^2 + y^2) 
- \delta_N  2\pi g |\psi|^2 \Big] \psi  \; , 
\eeq
where $g>0$ and the parameters 
$\delta_D=\pm 1$ and $\delta_N=\pm 1$, according 
as whether the axial kinetic term has anomalous 
($\delta_D =1$) or normal ($\delta_D=-1$) dispersion, 
and the nonlinear cubic term is self-focusing 
($\delta_N=1$) or self-defocusing ($\delta_N=-1$). 
In the context of nonlinear atom optics, the field 
$\psi({\bf r},t)$ of Eq. (1) 
is the macroscopic wave function of the 
Bose-Einstein condensate. In Eq. (1) the time $t$ is 
in units of $1/\omega_{\bot}$, where $\omega_{\bot}$ is the frequency 
of the transverse harmonic potential. The lengths 
are in units of the characteristic harmonic length 
$a_{\bot}=\sqrt{\hbar/(m\omega_{\bot})}$, where $m$ 
is the atomic mass. The mass $m$ is the true atomic mass 
in the absence of an axial confinement ($\delta_D=1$) 
\cite{r13,r14}, but it is instead the modulus of a negative 
effective mass which takes into account the effect of 
periodic axial potential when atoms are around the upper band 
edge ($\delta_D=-1$) \cite{r15,r16,r17}. 
The strength $g$ of the cubic nonlinearity 
is proportional to the modulus of the s-wave scattering 
length of the inter-atomic potential and its sign 
is given by the parameter $\delta_N$: 
$\delta_N=1$ corresponds to an effective interatomic attraction 
and $\delta_N=-1$ corresponds to an 
effective repulsion\cite{r13,r14,r15,r16}. 
\par 
In the context of nonlinear guided wave optics, 
the field $\psi({\bf r},t)$ of Eq. (1) represents 
instead the envelope 
of the electric field oscillating at a fixed 
frequency. Both the cubic nonlinearity and 
the transverse harmonic confinement model the 
refrective index of an inhomogeneous 
Kerr medium \cite{r17}. Note that in describing an optical 
pulse in fibers with Eq. (1) the two variables $t$ and $z$ 
have an inverted meaning: the variable $t$ is the axial 
coordinate of the electric field while the variable 
$z$ is the time coordinate of the electric field. 
The scaling of these variables is discussed, for instance,  
in Ref. \cite{r17}, where $g$ is set equal to $1/(2\pi)$ 
and the field $\psi({\bf r},t)$ is normalized 
such that $\int|\psi({\bf r},t)|^2 d^3{\bf r}$ 
represents the constant pulse energy. 
\par 
To study the 3D CNLSE time-dependent 
variational methods are often used 
\cite{r2,r10,r12,r18}. 
The 3D CNLSE is the Euler-Lagrange equation 
obtained by minimizing the following action functional 
\beq 
A = \int \psi^* 
\Big[ 
i{\partial \over \partial t} 
+{1\over 2} \nabla^2_{\bot} 
+ {\delta_D \over 2} {\partial^2 \over \partial z^2} 
- {1\over 2} (x^2 + y^2) + \delta_N \; \pi g |\psi|^2 \Big] 
\psi \; d^3{\bf r} \; dt \; . 
\eeq 
Some years ago we proposed \cite{r14} 
a variational ansatz of the field $\psi({\bf r},t)$ 
for the Eqs. (1) and (2) with $\delta_D=1$. 
The resulting effective non-polynomial Schr\"odinger 
equation is very accurate in reproducing 
the numerical results of the 3D CNLSE with $\delta_D=1$ 
\cite{r14,r19}. This effective equation has been 
used also to model the data 
of various experiments \cite{r20}. Here we use the same approach 
by considering also the case with $\delta_D=-1$. 
The variational ansatz is 
\beq 
\psi({\bf r},t) 
= f(z,t) 
{1 \over \pi^{1/2} \sigma(z,t)^2 } 
exp{\left( x^2+y^2 \over 2\sigma(z,t)^2 \right)}
\; , 
\eeq 
where $f(z,t)$ is the axial wave function and $\sigma(z,t)$ 
is the width of the transverse Gaussian wave function. 
By inserting Eq. (3) into Eq. (2) and 
integrating along $x$ and $y$, 
the resulting effective action functional 
depends on two fields: $f(z,t)$ and $\sigma(z,t)$. 
Neglecting the derivatives of $\sigma(z,t)$ we find 
that the Euler-Lagrange equation of the 
axial wavefunction $f(z,t)$ given by 
\beq 
i{\partial f \over \partial t} = 
\Big[ 
-{\delta_D \over 2} {\partial^2 \over \partial z^2} 
-\delta_N {g \over \sigma^2} |f|^2 
+ {1\over 2}\left( {1\over \sigma^2} + \sigma^2 \right)
\Big] f \; , 
\eeq 
and the Euler-Lagrange equation of the transverse 
width $\sigma(z,t)$ is instead given by 
\beq 
\sigma^2=\sqrt{1 - \delta_N \; g |f|^2} \; . 
\eeq 
Inserting this formula into Eq. (4), 
we obtain a nonpolynomial Schrodinger equation (NPSE). 
In the weak-coupling limit $g|f|^2 \ll 1$, 
where $\sigma \simeq 1$, the NPSE reduces to the familiar 
1D cubic nonlinear Schr\"odinger equation (1D CNLSE) 
\beq 
i{\partial f \over \partial t} =
\left[
-{\delta_D \over 2} {\partial^2 \over \partial z^2} 
- \delta_N \; g |f|^2 \right] f \; ,  
\eeq 
where the additive constant $1$ has been omitted 
because it does not affect the dynamics. 
\par 
It is well known that the 1D CNLSE admits solitonic 
solutions. In particular, for $\delta_D \cdot \delta_N =1$, 
i.e. for anomalous dispersion ($\delta_D=1$) and 
self-focusing nonlinearity ($\delta_N=1$) 
but also for normal dispersion ($\delta_D =-1$) 
and self-defocusing nonlinearity ($\delta_N=-1$), 
one finds the localized bright soliton solution 
\beq 
f(z,t) = \sqrt{g \over 4} \; 
Sech\left[ {g (z - v t) \over 2}\right] 
e^{i \delta_D v (z - v t )} e^{i (v^2/(2\delta_D) - \mu )t}   \; , 
\eeq 
where $Sech[x]$ is the hyperbolic secant, 
$v$ is the arbitrary velocity 
of propagation of the shape-invariant bright soliton and 
the parameter $\mu = - \delta_D {g^2/8}$ is obtained 
by the normalization 
condition of the field $f(z,t)$ to one \cite{r11,r12}. 
Instead, for $\delta_D \cdot \delta_N =-1$, 
i.e. for normal dispersion ($\delta_D=-1$) and 
self-focusing nonlinearity ($\delta_N=1$) 
but also for anomalous dispersion ($\delta_D =1$) 
and self-defocusing nonlinearity ($\delta_N=-1$), 
one finds the localized but 
not normalized dark soliton solution 
\beq 
f(z,t) = \sqrt{g\phi_{\infty}^2 -v^2\over g}  \; 
\Big( 
Tanh\left[(z - v t)\sqrt{g\phi_{\infty}^2-v^2} \right] 
+ i {v\; \delta_D \over \sqrt{g\phi_{\infty}^2-v^2}}\Big) e^{-i\mu t}  \; ,
\eeq 
where $Tanh[x]$ is the hyperbolic tangent, 
$\phi_{\infty}>0$ is the constant value of the field 
amplitude $|f(z,t)|$ 
at infinity ($z\to \pm \infty$), and 
$v$ is the velocity of propagation of the dark soliton. 
The parameter $\mu =\delta_D g\phi_{\infty}^2$ is obtained 
by the asymptotic behavior of the field $f(z,t)$. 
For $0<|v|<c_s$, where $c_s=\sqrt{g\phi_{\infty}^2}$ is the sound 
velocity, the minimum value of the field $f(z,t)$ 
is greater than zero and the wave 
is called gray soliton. If $v=0$ (stationary 
dark soliton) the minimum of the field $f(z,t)$ is zero and 
the wave is called black soliton \cite{r11,r12}. 
The velocity $c_s$ is the maximal velocity of the dark 
soliton, namely the velocity of a wave 
of infinitesimal amplitude, i.e. a sound wave, 
propagating in a medium of density 
$\rho_{\infty}=\phi_{\infty}^2$. 

\section{3D bright and dark solitons} 

We have seen that for the 1D CNLSE 
the axial density profiles of bright and dark solitons 
do not depend on the sign of $\delta_D$ and $\delta_N$, 
but only by their product 
$\delta_D\cdot \delta_N$ which discriminates 
between bright and dark solitons. 
It is important to stess that this property 
is valid only in the 1D limit. In fact, 
as previously shown, the 1D CNLSE is only the 
weak-coupling limit of 1D NPSE, 
that is derived from the 3D CNLSE. 
In the remaining part of the paper we show that, 
in general, for a fixed value of the strength $g$, 
3D bright solitons ($\delta_D \cdot \delta_N=1$) 
with anomalous dispersion ($\delta_D =1$) 
and self-focusing nonlinearty ($\delta_N=1$) 
do not have the same axial density profile of 
bright solitons with normal dispersion ($\delta_D=-1$) 
and self-defocusing nonlinearity ($\delta_N=-1$). 
The same happens for 3D dark solitons. 
In the remaining part of the paper we call 3D solitons 
the solitary solutions of the NPSE, which are 
good approximations of the exact solutions of the 3D CNLSE 
under transverse confinement. 
\par 
To analyze 3D bright and dark solitons 
we start from the NPSE and set 
\beq 
f(z,t) = \phi(z-vt) 
e^{i \theta(z-vt)} e^{i (v^2/(2\delta_D) - \mu )t} \; , 
\eeq 
where $\zeta = z-vt$ is the comoving coordinate 
of the soliton and both $\phi(\zeta )$ 
and $\theta(\zeta )$ are real fields. 
In this way we get an equation for $\phi(\zeta )$, namely 
\beq 
\left( \mu - {v^2\over 2\delta_D} + v \theta' \right) \phi = 
-{\delta_D \over 2} \left( \phi'' - \theta'^2 \phi \right) 
+ {\phi - \delta_N (3/2) g \phi^3 \over 
\sqrt{1 - \delta_N\; g \phi^2} } 
\; , 
\eeq
and also an equation for the phase $\theta( \zeta )$, 
that is 
\beq 
v \; \phi' = {\delta_D \over 2}
\left( \phi \; \theta'' + 2 \phi' \theta' \right) 
\; . 
\eeq 
Note that the prime means the derivative with 
respect to $\zeta$; moreover, if $\phi = 0$ then 
from Eq. (11) one finds $\theta'= v/\delta_D$. 

\begin{figure}
\centerline{\includegraphics[width=10cm]{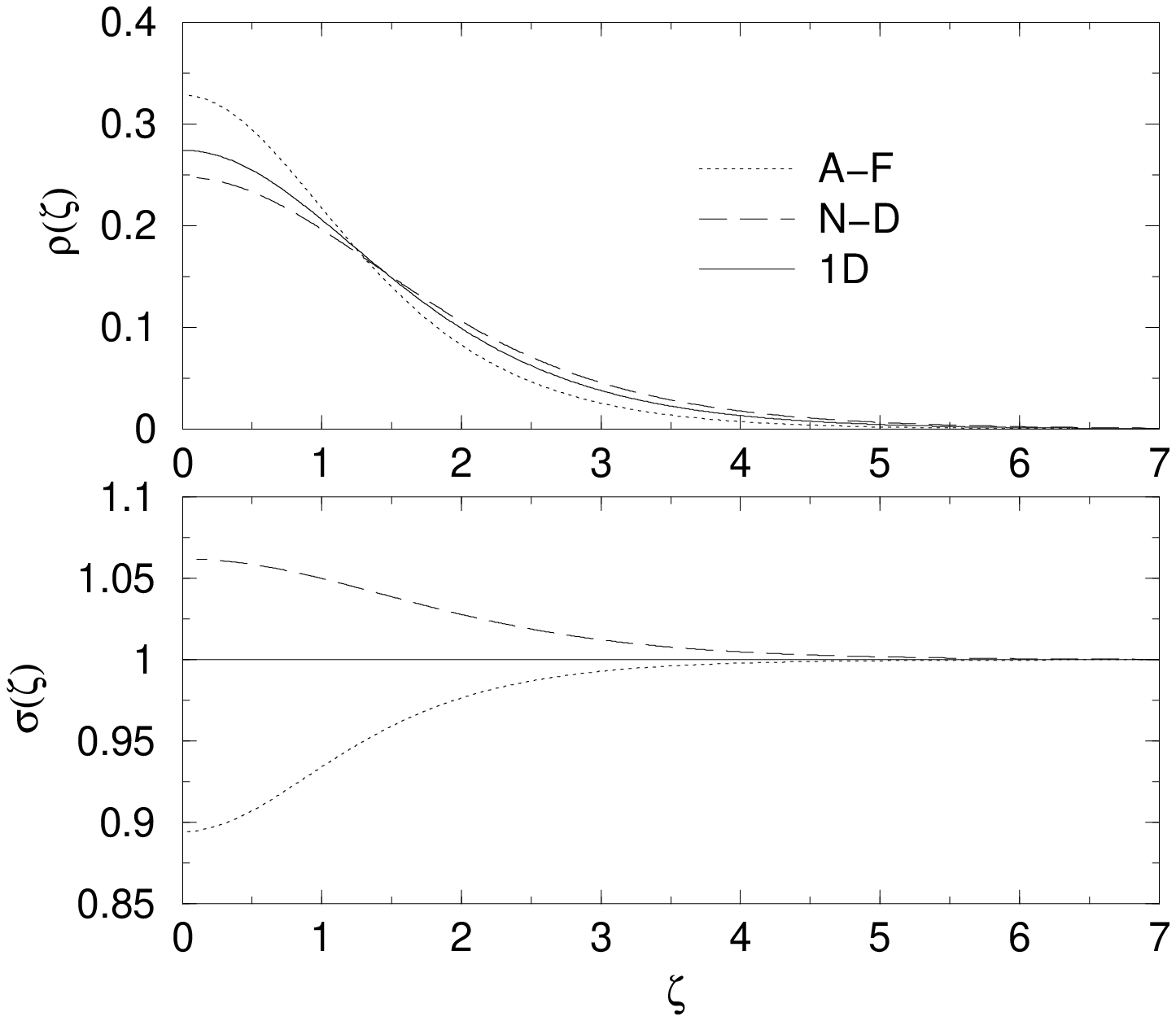}} 
{FIG. 1. Axial density profile $\rho(\zeta)=\phi(\zeta )^2$
and transverse width $\sigma(\zeta )$ of the
bright soliton ($\delta_D\cdot \delta_N=1$).
Strength of the nonlinearity: $g=1$.
A-F means 3D bright soliton
with anomalous dispersion ($\delta_D=1$) and self-focusing
nonlinearity ($\delta_N=1$); N-D means
3D bright soliton with normal dispersion
($\delta_D=-1$) and self-defocusing nonlinearity ($\delta_N=-1$).}
\end{figure} 

The two equations (10) and (11) are coupled but the second one 
can be written, for $\phi \neq 0$, as 
\beq 
v {\left( \phi^2 \right)' \over \phi} 
= 
\delta_D {\left( \phi^2 \theta' \right)' \over \phi } \; , 
\eeq 
from which we find 
\beq 
v \; \phi^2 = \delta_D \; \phi^2 \theta' + \xi \; , 
\eeq
where the integration constant $\xi =0$ for bright solitons 
and $\xi =v\phi_{\infty}^2$ for dark solitons, 
with $\phi_{\infty}$ the value of the field 
$\phi(\zeta )$ at $\zeta = \pm \infty$. 
Because $\phi_{\infty}=0$ for bright solitons, 
the Eq. (13) can be expressed as 
\beq 
\theta' = {v \over \delta_D} 
\left( 1-{\phi_{\infty}^2 \over \phi^2 } \right) \; .  
\eeq 
By using this formula, 
the equation of $\phi(\zeta )$ becomes 
\beq 
\phi'' = - {\partial V\over \partial \phi } \; , 
\eeq 
where 
\beq 
V(\phi ) = {1 \over \delta_D } 
\left[ 
\mu \; \phi^2 - \phi^2 \sqrt{1 - \delta_N\; g \phi^2} 
+ {v^2 \phi_{\infty}^4 \over \delta_D}{1 \over \phi^2}
\right] \; . 
\eeq
Thus, the field $\phi(\zeta )$ can be thought as 
the coordinate $\phi$ 
of a fictitious particle at time $\zeta$. 
In this picture $V(\phi )$ is the external 
potential acting on the fictitious particle. 
The constant of motion is then given by 
\beq
K = {1 \over 2} \phi'^2 + V(\phi)
\eeq 
and after separation of variables we get  
\beq 
d\zeta = {d\phi \over 
\sqrt{2\left(K - V(\phi) \right)} } \; . 
\eeq 
We stress that in our previous papers [13b] and \cite{r14} 
only anomalous dispersion ($\delta_D=1$) 
and $v\, \phi_{\infty}^2=0$ were considered. 
Here we investigate for the first time 
both normal dispersion ($\delta_D=-1$) and 
$v \, \phi_{\infty}^2\neq 0$. In particular, 
the Eq. (14) and the centrifugal term of Eq. (16) are new 
and essential to study the case $v \, \phi_{\infty}^2 \neq 0$, 
i.e gray solitons with normal or anomalous dispersion. 

\subsection{Bright solitons} 

The NPSE admits bright solitons for 
$\delta_D\cdot \delta_N =1$. In this case we have 
$\phi_{\infty} =0$ and $K=0$. By using the Eq. (18) 
we find that the bright soliton $\phi(\zeta )$ 
satisfies the implicit formula 
$$ 
\zeta = {1\over \sqrt{2}} \sqrt{\delta_D \over 1-\mu} 
ArcTanh\left[\sqrt{ \sqrt{1-\delta_N \; g \phi^2}
-\mu\over 1-\mu} \right]  
$$
\beq 
- {1\over \sqrt{2}} \sqrt{\delta_D \over 1+\mu}
ArcTan\left[\sqrt{ \sqrt{1-\delta_N \; 
g \phi^2}-\mu\over 1+\mu} \right] 
\; ,  
\eeq 
where $ArcTanh[x]$ and $ArcTan[x]$ are the 
hyperbolic arctangent and the arctangent, 
respectively. They satisfy the formula $ArcTan[i x]=i \;ArcTanh[x]$ 
which is useful when $\delta_D=-1$. 
It is not difficult to show that this equation reduces 
to Eq. (7) in the 1D weak-coupling limit $g\phi^2\ll 1$ taking 
into account the relationship between $\mu$ 
and $g$. This relationship is obtained by imposing 
the normalization condition to the bright soliton. 
Choosing $\int \phi(\zeta )^2 d\zeta =1$ we obtain 
\beq 
g = {2\sqrt{2} \over 3} (2\mu + 1) 
\sqrt{1-\mu \over \delta_D} \; . 
\eeq 
It is clear that for $\delta_D=-1$ it must be $\mu >1$, 
while for $\delta_D=1$ one has $0<\mu <1$ and the 
3D bright soliton is dynamically stable only 
for $1/2 <\mu <1$ \cite{r13,r14}. 
\par
In Fig. 1 we plot the bright soliton 
with nonlinearity $g=1$. 
The figure shows that the 1D bright soliton 
given by Eq. (7) and the two 3D bright solitons 
given by Eq. (18) have different axial profiles 
because their transverse profiles are quite different. 
For the 1D bright soliton $\sigma(\zeta ) =1$, 
while for the two 3D bright solitons $\sigma(\zeta )$ is not
constant and depends on the sign $\delta_N$ of the
nonlinearity: $\sigma(\zeta) \ge 1$ with $\delta_N=-1$
(self-defocusing and normal dispersion) 
and $\sigma(\zeta )\le 1$ with $\delta_N=1$
(self-focusing and anomalous dispersion). 
\par 
In Fig. 2 we plot the axial width $z_{F}$ and 
transverse width $\sigma(0)$ of the bright soliton 
as a function of the strength $g$. 
As expected, the axial width decreases by increasing $g$ 
but for the A-F bright soliton, i.e. the 3D bright soliton 
with anomalous dispersion ($\delta_D=1$) 
and self-focusing ($\delta_N=1$), there is a 
critical strength $g_c=4/3=1.3\bar3$ above which 
the soliton does not exist anymore (see also \cite{r5}). 
This critical value is in excellent 
agreeent with the numerical result $g_c=1.35$ 
obtained by solving the 3D CNLSE \cite{r21}. 
\par 
For the N-D bright soliton, i.e. the 3D bright soliton 
with normal dispersion ($\delta_D=-1$) 
and self-defocusing ($\delta_N=-1$), there is not 
a critical strength and the N-D bright soliton 
exist for any $g$, as the strictly 1D bright soliton. 

\begin{figure}
\centerline{\includegraphics[width=10cm]{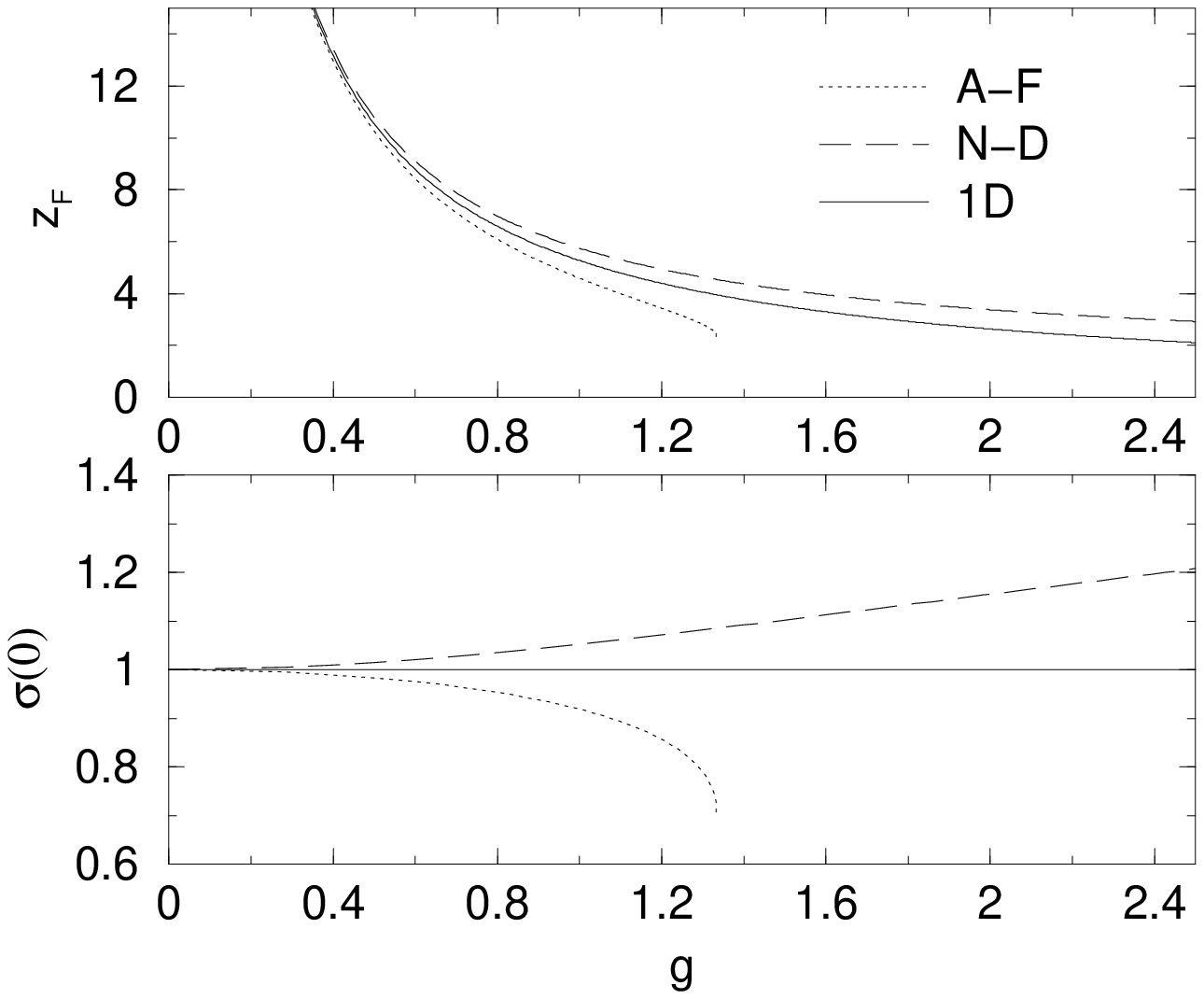}}
{FIG. 2. Axial full width half maximum $z_{F}$ and 
transverse width $\sigma$ at $\zeta =0$ as a function 
of the strength $g$ of the nonlinearity for the 
bright soliton ($\delta_D\cdot \delta_N=1$). 
The meaning of labels is the same of Fig. 1.} 
\end{figure}

The lower panel of Fig. 2 shows that the transverse 
width of the N-D bright soliton grows indefinitely 
and a remarkable consequence is that the soliton geometry changes 
from cigar-shaped to disk-shaped by increasing 
the strength $g$ of the cubic nonlinearity: 
at $g = 11.94$ we find $z_{F}=\sigma $. 

\subsection{Black and gray solitons} 

The NPSE admits dark solitons for 
$\delta_D\cdot \delta_N = -1$. In this case we have 
$\phi_{\infty} \neq 0$ and $K=V(\phi_{\infty})$. 
By using the Eq. (17) we first numerically determine 
the dark soliton $\phi(\zeta )$ with $v=0$. 
As previously stressed this dark soliton is called 
black soliton because its minimum value is zero. 
For the black soliton from Eq. (11) one finds $\theta'=0$. 

\begin{figure}
\centerline{\includegraphics[width=10cm]{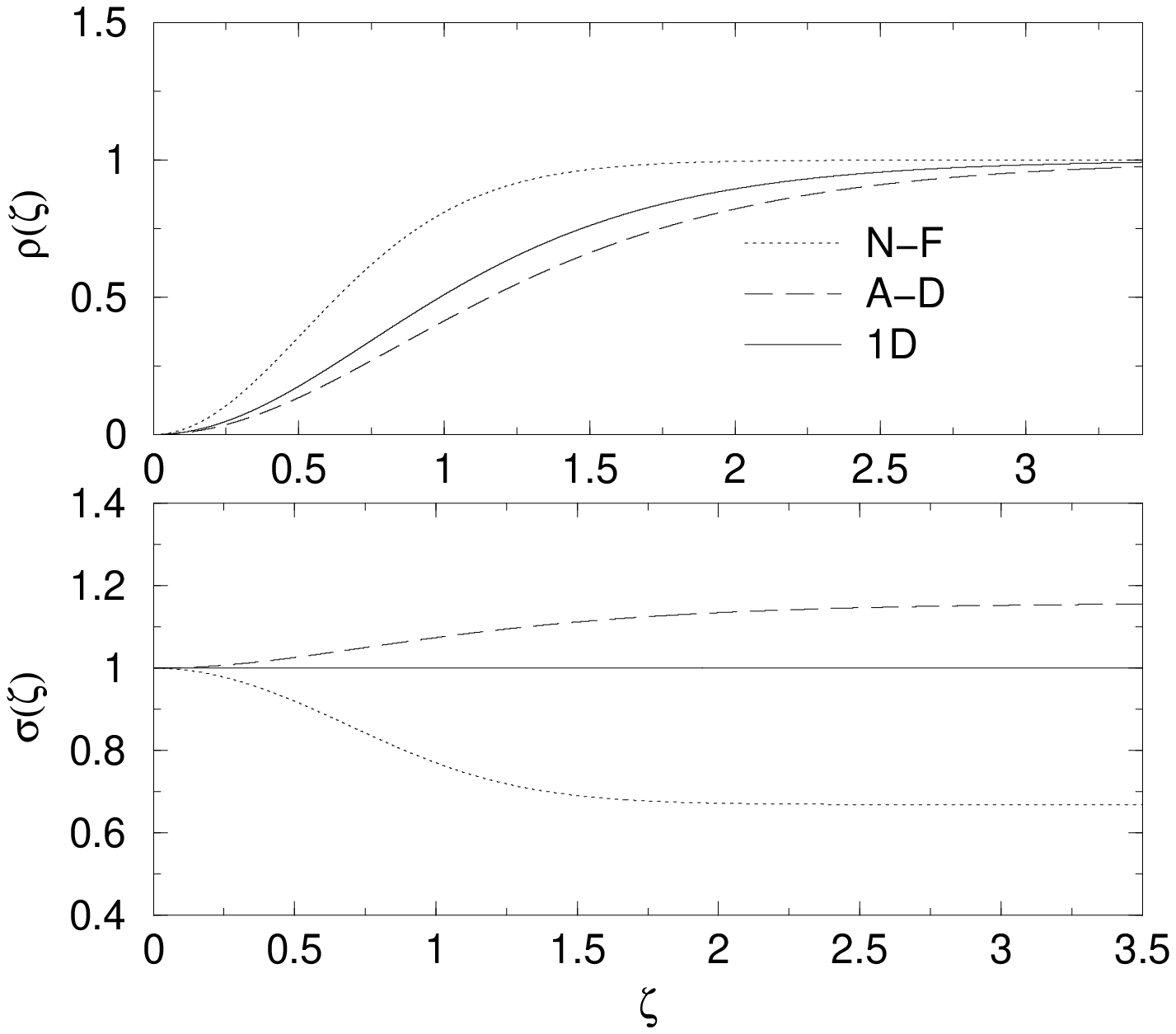}}
{FIG. 3. Axial density profile $\rho(\zeta)=\phi(\zeta )^2$
and transverse width $\sigma(\zeta )$ of the
dark soliton ($\delta_D\cdot \delta_N=-1$) with $v=0$
(black soliton). Strength of the nonlinearty $g=1$
and asymptotic field $\phi_{\infty}=1$.
A-D means 3D black soliton
with anomalous dispersion ($\delta_D=1$) and self-defocusing
nonlinearity ($\delta_N=-1$); N-F means 3D
black soliton with normal dispersion ($\delta_D=-1$) and
self-focusing nonlinearity ($\delta_N=1$).}
\end{figure}

In Fig. 3 we plot the axial profile $\rho(\zeta)=
\phi(\zeta )^2$ and the transverse width $\sigma(\zeta )$
of the black soliton setting $g=1$ and $\phi_{\infty}=1$. 
We compare the 1D black soliton given by Eq. (8) and the 
3D black solitons obtained by numerically solving 
Eq. (15) with Eq. (16). The transverse width 
is $\sigma\simeq 1$ near the black hole but 
for 3D dark solitons it is 
$\sigma = (1-\delta_N \phi_{\infty}^2)^{1/4}$ at infinity. 
It is clear that this condition bounds the domain 
of existence of the 3D dark soliton with 
self-focusing nonlinearity ($\delta_N=1$): 
this dark soliton exist only for $g\phi_{\infty}^2<1$. 
\par
When $v\neq 0$ then the dark soliton is called 
gray soliton and its minimum value is greater than zero.  
In Fig. 4 we plot the hole of the 3D gray soliton 
with anomalous dispersion ($\delta_D=1$) and 
self-defocusing nonlinearity ($\delta_N=1$) for some values of $v$. 
By increasing the velocity $v$ 
the depth of the hole is reduced and becomes 
zero when the velocity reaches the sound velocty $c_s$, 
given in general by the formula 
\beq 
c_s = 
\sqrt{ - {5\over 4} 
{\delta_N g \rho_{\infty} 
\over \sqrt{1-\delta_N g \rho_{\infty}} 
}
+ {1\over 4} 
{\delta_N g \rho_{\infty} - 2\delta_N^2 g^2 \rho_{\infty}^2 
\over 
\left( 1 - \delta_N g \rho_{\infty} 
\right)^{3/2} 
}
}
\; , 
\eeq 
where $\rho_{\infty}=\phi_{\infty}^2$ \cite{r22}. 
The 3D gray soliton with $\delta_N=-1$ exist for any $g$ 
if $v$ is sufficiently small. Nevertheless, 
it has been shown \cite{r23} 
that these 3D dark solitons become 
dynamically unstable for a large nonlinearity $g$ 
($g=1.5$ for the black soliton) due to the so-called 
snake instability. The snake instability implies the apparence of 
a non-axisymmetric purely imaginary eigenvalue in the Bogoliubov 
spectrum of elementary excitations. In this case the black soliton decays 
into a solitonic vortex \cite{r24}. 

\begin{figure}
\centerline{\includegraphics[width=10cm]{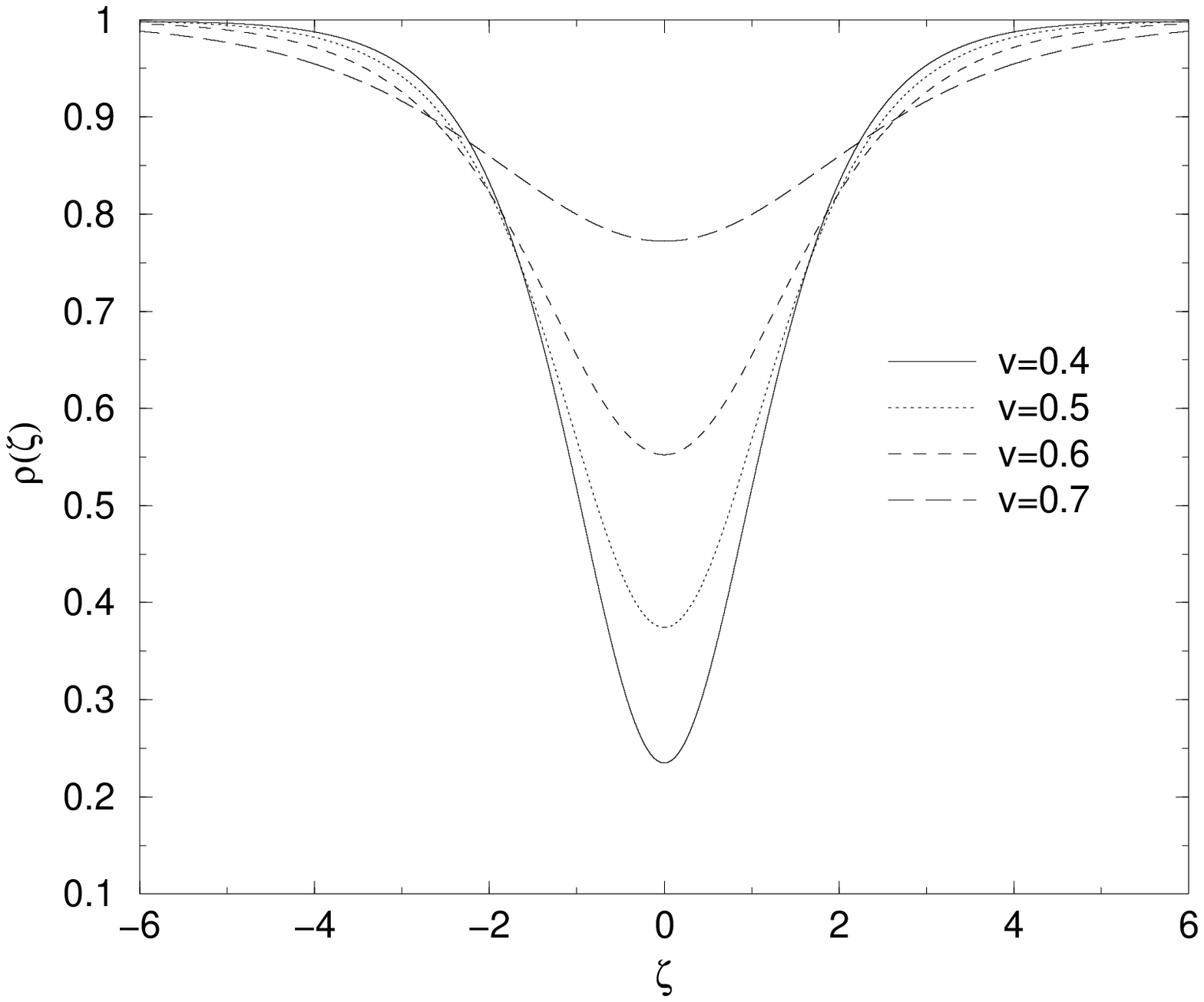}}
{FIG. 4. Axial density profile $\rho(\zeta)=\phi(\zeta )^2$ 
of the gray soliton ($v\ne 0$) with anomalous 
dispersion ($\delta_D=1$) and self-defocusing nonlinearity ($\delta_N=-1$). 
Strength of the nonlinearty $g=1$ and asymptotic field 
$\phi_{\infty}=1$.  }
\end{figure} 

In addition, it has been shown that many complex modes appear 
in the 3D black soliton at $g=4$; for this nonlinearity 
an axisymmetric vortex ring 
emerges with energy intermediate between that of 
the black soliton and the solitonic vortex \cite{r24}. 

\section{Conclusions} 

We have studied 3D bright and dark 
solitons with anomalous
and normal dispersion taking into account their transverse
width by using a non-polynomial Schr\"odinger equation. 
We have found new general analytical solutions 
for 3D bright solitons under transverse harmonic confinement. 
In addition, we have also analyzed 3D black and gray 
solitons showing that self-focusing dark solitons 
(and with normal dispersion) exist only below 
a critical axial density. As a by-product we have 
obtained a formula for the sound velocity, 
i.e. the maximal velocity of propagation of a gray soliton, 
useful for both self-focusing and self-defocusing media. 
Atomic matter waves of Bose-Einstein condensates 
are now rutinely produced with dilute gases 
of alkali-metal atoms at ultralow temperatures. 
Also optical pulses in graded-index fibers 
are now available. As previously stressed, our predictions 
on 3D solitons under transverse confinement 
can be experimentally tested using these materials.


\section*{References} 

\begin{thebibliography}{9}

\bibitem{r1} C.J. Pethick and H. Smith, {\it 
Bose-Einstein Condensation in Dilute Gases} 
(Cambridge University Press, Cambridge, 2001); 
L.P. Pitaevskii and S. Stringari, {\it Bose-Einstein Condensation} 
(Clarendon Press, Oxford, 2003). 

\bibitem{r2} G.P. Agrawal, {\it Nonlinear Fiber Optics} 
(Academic Press, San Diego, 1995). 

\bibitem{r3} L. Deng, E. Hagley, J. Wen, M. Trippenbach, 
Y. Band, P. Julienne, J. Simsarian, K. Helmerson, S. Rolston,  
and W.D. Phillips, Nature {\bf 398}, 218 (1999).  

\bibitem{r4} E.W. Hagley, L. Deng, M. Kozuma, J. Wen, K. Helmerson, 
S. Rolston, and W.D. Phillips, Science {\bf 283}, 1706 (1999). 

\bibitem{r5} S. Burger, K. Bongs, S. Dettmer, W. Ertmer, K. Sengstock, 
A. Sanpera, G.V. Shlyapnikov, and M. Lewenstein, 
Phys. Rev. Lett. {\bf 83}, 5198 (1999).  

\bibitem{r6} J. Denschlag, J.E. Simsarian, 
D.L. Feder, C.W. Clark, L.A. Collins, 
J. Cubizolles, L. Deng, E.W. Hagley, K. Helmerson, W.P. Reinhardt, 
S.L. Rolston, B.I. Schneider, and 
W.D. Phillips, Science {\bf 287}, 97 (2000). 

\bibitem{r7} K.E. Strecker, G.B. Partridge, A.G. Truscott, and R.G. Hulet, 
Nature {\bf 417}, 150 (2002). 

\bibitem{r8} L. Khaykovich, F. Schreck, G. Ferrari, T. Bourdel, 
J. Cubizolles, L.D. Carr, Y. Castin Y, and C. Salomon, 
Science {\bf 296}, 1290 (2002). 

\bibitem{r9} B. Eiermann, Th. Anker, M. Albiez, M. Taglieber, 
P. Treutlein, K.-P. Marzlin, and M.K. Oberthaler, Phys. 
Rev. Lett. {\bf 92}, 230401 (2004). 

\bibitem{r10} A. Hasegawa and Y. Kodama, 
{\it Solitons in Optical Communications} 
(Clarendon Press, Oxford, 1995); 
N.N. Akhmediev and A. Ankiewicz, 
{\it Solitons: Nonlinear Pluses and Beams} 
(Chapman and Hall, New York, 1997). 

\bibitem{r11} V.E. Zakharov and A.B. Shabat, 
Soviet Physics JETP {\bf 34} 62 (1972). 

\bibitem{r12} M.J. Ablowitz and H. Segur, 
{\it Solitons and the Inverse Scattering Transform} 
(SIAM, Philadelphia, 1981); S.P. Novikov, S.V. Manakov, 
L.P. Pitaevskii, and V.E. Zakharov  
{\it Theory of Solitons. The Inverse Scattering Method} 
(Plenum Press, New York, 1984); C. Sulem and P. Sulem, 
{\it The Nonlinear Schr\"odinger Equation} (Springer, New-York, 1999).  

\bibitem{r13} (a) V.M. Perez-Garcia, H. Michinel, and 
H. Herrero, Phys. Rev. A {\bf 57}, 3837 (1998); (b) 
L. Salasnich, A. Parola, and L. Reatto, 
Phys. Rev. A {\bf 66}, 043603 (2002). 

\bibitem{r14} L. Salasnich, A. Parola, and L. Reatto, 
Phys. Rev. A {\bf 65}, 043614 (2002). 

\bibitem{r15} V. A. Brazhnyi and V. V. Konotop, 
Mod. Phys. Lett. B {\bf 18}, 627 (2004);  
D. E. Pelinovsky, A. A. Sukhorukov, and Y. S. 
Kivshar, Phys. Rev. E {\bf 70}, 036618 (2004). 

\bibitem{r16} K.M. Hilligsoe, M.K. Oberthaler, and 
K.P. Marzlin, Phys. Rev. A {\bf 66}, 063605 (2002).

\bibitem{r17} S. Raghavan and G.P. Agrawal, 
Opt. Commun. {\bf 180}, 377 (2000); J. Jasinski, 
Opto-Electron. Rev. {\bf 13}(2), 129 (2005). 

\bibitem{r18} E. Cerboneschi, R. Mannella, E. Arimondo E, and 
L. Salasnich, Phys. Lett. A {\bf 249}, 495 (1998);  
L. Salasnich, Int. J. Mod. Phys. B {\bf 14}, 1 (2000); 
L. Salasnich, A. Parola and L. Reatto, 
J. Phys. B: At. Mol. Opt. {\bf 35}, 3205 (2002).  

\bibitem{r19} M. Modugno, C. Tozzo, and F. Dalfovo, 
Phys. Rev. A {\bf 71}, 019904 (2005); 
C. Tozzo, M. Kramer and F. Dalfovo, 
Phys. Rev. A {\bf 72}, 023613 (2005). 

\bibitem{r20} See for instance the paper \cite{r9}. 

\bibitem{r21} A. Gammal, L. Tomio, and 
T. Frederico, Phys. Rev. A {\bf 66}, 043619 (2002). 

\bibitem{r22} This formula extends that 
found for the self-defocusing case ($\delta_N=-1$) 
in L. Salasnich, A. Parola and 
L. Reatto, Phys. Rev. A. {\bf 69}, 045601 (2004). 

\bibitem{r23} A.E. Muryshev, 
H.B. van Linden van den Heuvell, 
and G. V. Shlyapnikov, Phys. Rev. A {\bf 60}, R2665 (1999); 
D.L. Feder, M.S. Pindzola, L.A. Collins, 
B.I. Schneider, and C.W. Clark, 
Phys. Rev. A {\bf 62}, 053606 (2000). 

\bibitem{r24} 
S. Komineas and N. Papanicolaou, 
Phys. Rev. Lett. {\bf 89}, 070402 (2002); 
S. Komineas and N. Papanicolaou, 
Phys. Rev. A {\bf 67}, 023615 (2004); 
S. Komineas and N. Papanicolaou, 
Phys. Rev. A {\bf 68}, 043617 (2003); 
S. Komineas and N. Papanicolaou, 
Laser Phys. {\bf 14}, 571 (2004). 

\end{thebibliography}
\end{document}